\documentclass[11pt]{article}

\usepackage{amsmath,amssymb,amsthm,bm,enumerate,mathrsfs,mathtools}
\usepackage{amsfonts}
\usepackage{framed}
\usepackage{hyperref}
\usepackage{bm}
\usepackage{soul}
\usepackage{graphicx}
\usepackage{comment}
\usepackage{enumitem}
\usepackage{titlesec}
\usepackage{changepage}

\usepackage[format=plain,
            labelfont=it,
            textfont=it]{caption}

\newtheorem{thm}{Theorem}

\newtheorem{remark}[thm]{Remark}

\usepackage{authblk}

\begin{document}

\title{\bf Inference after black box selection}

\author[1]{Jelena Markovic}
\author[1]{Jonathan Taylor}
\author[2]{Jeremy Taylor}
\affil[1]{Department of Statistics, Stanford University}
\affil[2]{Via Science}
\date{}

\maketitle

\begin{abstract}
	We consider the problem of inference for parameters selected to report only after some algorithm, the canonical example being inference for model parameters after a model selection procedure. The conditional correction for selection requires knowledge of how the selection is affected by changes in the underlying data, and current research explicitly describes this selection. In this work, we assume 1) we have \textit{in silico} access to the selection algorithm and 2) for parameters of interest, the data input into the algorithm satisfies (pre-selection) a central limit theorem jointly with an estimator of our parameter of interest. Under these assumptions, we recast the problem into a statistical learning problem which can be fit with off-the-shelf models for binary regression. The feature points in this problem are set by the user, opening up the possibility of active learning methods for computationally expensive selection algorithms. We consider two examples previously out of reach of this conditional approach: stability selection and multiple cross-validation. %In terms of dimensions of problem, for moderate computational costs we can address problems in dimensions in the hundreds, higher than those of POSI, our closest competitor in terms of generality of selection. 
\end{abstract}

\begin{comment}
\begin{abstract}
Given a variable selection algorithm, practitioners tend to re-ran the algorithm on random subsamples of the data to access the robustness of the selection. In other cases, where the model selection procedures are randomized, practitioners tends to re-ran the algorithm for different randomization instances. The multiple models are then combined into one containing the most stable variables -- all the variables selected above a certain number of times. 
An important question becomes how to evaluate significance of the variables in the final set given that the set has been selected based on data via a complex model selection procedure. In order to provide valid inference post-selection, one needs to account for (randomized) selection by conditioning on the observed variables. The event of selecting each of these variables is usually hard to describe analytically, thus we turn to a statistical learning algorithm to estimate the selection probabilities as a function of data. Assuming we have access to the possibly randomized selection algorithm, we re-ran the algorithm on the perturbed data learning the selection probabilities as a function of data. 
The proposed method solves a selective inference problem with general applicability to any black box selection algorithms by casting it in an estimation framework.
\end{abstract}
\end{comment}

%---------------------------------------------------
\section{Introduction}
%---------------------------------------------------

In the era of big data and high dimensional statistics, practitioners usually run various algorithms on their data to find interesting questions to ask. Statistically, this means they use exploratory tools to form data-dependent hypotheses. As model selection procedures present one of the most useful tools for reducing data dimensionality and understanding and interpreting a given dataset, practitioners tend to form hypotheses based on a data-dependent set of variables obtained via a model selection procedure.

After running a complicated algorithm to choose the parameters to test, \ul{the question is how to assess the significance of these parameters post-selection, i.e.~using the same dataset.} Classical $p$-values and confidence intervals for the selected parameters computed on the same data used for selection are not longer valid due to \textit{data snooping} issues, i.e.~forming and testing random hypotheses using the same data \cite{berk2013valid}. The conditional approach to selective inference methodology solves this problem by modifying/correcting the normal density of a test statistic with the appropriate \textit{selection probabilities given data}, a crucial object needed for selective inference that depends on the selection procedure. In case there is a simple analytic description of the selection event in terms of data and possible randomization, this selection probability function (of data) is easy to compute.
The inference methods developed in the literature heavily rely on the tractability of selection to conduct valid inference after various model selection algorithms including Lasso, forward-stepwise regression \cite{lee2013exact, sequential_post_selection} and their randomized counter-parts \cite{selective_sampler}. 
For more complicated selection algorithms, however, the description of the selection region might not have a simple analytic form. In these cases the selection probabilities given data (conditional corrections) are hard to compute explicitly or even approximate.

In this work, \ul{we propose a new selective inference method for any possibly randomized black box algorithm.} \ul{We develop a learning method for estimating these probabilities by perturbing the original data along a line and re-running the whole model selection black box algorithm on the perturbed data.} For each perturbed data sample, we only need to record whether the originally observed selection event occurred on the perturbed sample. Perturbed data and their corresponding selection binary indicators comprise a newly generated dataset with one-dimensional covariates and binary response. The selection probabilities as a function of data represent the probabilities of success (or observing positive outcome) given covariate value for the binary response in this generated dataset. We propose several estimation procedures for fitting this data. Having estimated the selection probabilities as a function of data, we easily get $p$-values and confidence intervals for data-dependent parameters.

We apply our method to pretty intractable model selection procedures that are developed as practical improvements over the existing procedures.
Even though plenty of algorithms are available for variable selection with good theoretical properties, applying them to real datasets is usually challenging. Most of these algorithms have tuning parameters that are notoriously hard to choose well in practice. In order to get a more convincing selected set of variables, practitioners re-run a given algorithm to improve on the original procedure.
Given a randomized model selection procedure, statisticians often re-run the procedure for different randomization samples to access the stability and robustness of the procedure. In the case of non-randomized model selection procedure, statisticians re-ran the selection algorithm on the random data subsamples for the same reason. Seeing the same variables appear many times brings more confidence into the variables being truly associated with the response. On the other hand, seeing some variables appear only a few times indicates that they might not be as important.

Popular examples include:
\begin{itemize}[leftmargin=*]
\item \textbf{Stability selection} algorithm introduced in \cite{meinshausen2010stability} runs the whole Lasso path for many random subsamples of the data. Based on the many selected models, one for each subsample and each penalty level, the authors aggregate them to get a single final selected set (more details later). A similar method can be applied to many other model selection procedures to improve and enhance the underlying algorithms. With stability selection, the results are much less sensitive to the choice of the regularization. Furthermore, the authors show that stability selection makes variable selection consistent in settings where the original methods fail.

\item \textbf{Multiple cross-validation} is a model selection algorithm that provides a more robust version of the standard cross-validation (CV). Applied to the Lasso algorithm, standard CV chooses a penalty level that depends on the data and also on a random choice of the folds/splits of the data. A CV procedure then runs the Lasso on the whole dataset with the cross-validated penalty level to get a selected set of variables.
 Multiple CV runs the CV procedure multiple times on random data subsamples in order to assess the variability of the selected sets. The multiple selected sets (one for each CV run) are then aggregated to consist of variables selected above a given number of times.

%\item \textbf{Model-$\bm X$ knockoffs} of \cite{candes2016panning} is a model selection procedure procedure where the added randomization is injected into the selection algorithm. Even though the selected set of variables for a given randomization sample controls False Discovery Rate (FDR), it is natural in practice to repeat the procedure multiples times. Even though the final set of variables does not have FDR guarantees, the authors of \cite{candes2016panning} still re-ran the procedure multiple times in their application. The final set of variables are the ones selected above a certain number of times.
\end{itemize}

%---------------------------------------------------
\subsection{Outline}
%---------------------------------------------------

The structure of the paper is as follows.
In Section \ref{sec:simple:example}, we present a simple problem, solving one-dimensional selective inference problem with the explicit formula for the selection probability given data. In Section \ref{sec:regression}, we introduce the general conditional inference framework for the regression problem. In Section \ref{sec:learning}, we discuss how to learn the conditional selection probabilities given data, enabling valid post-selection inference after complex model selection procedures. In Section \ref{sec:simple:example:revisited}, we explain how the proposed estimation procedure looks like in the case of the simple example.
In Section \ref{sec:stability:selection} and Section \ref{sec:cv}, we apply the proposed selective inference method to stability selection and multiple CV procedure, respectively. Section \ref{sec:discussion} discusses potential extensions of this approach and interesting future research.

%%%%%%%%%%%%%%%%%%%%%%%%%%%%%%%%%%%%%%%%%%%%%%%%%%%%
%%%%%%%%%%%%%%%%%%%%%%%%%%%%%%%%%%%%%%%%%%%%%%%%%%%%

%-------------------------------------------------
\section{Simple example} \label{sec:simple:example}
%-------------------------------------------------

Before presenting the proposed method in a more general regression setting, we first focus on a simple example, which contains the introductory ideas, making the exposition easier. The problem here is to estimate the mean parameter after multiple randomized selections. Assume the data consists of $n$ samples $Y_1,\ldots, Y_n\overset{i.i.d.}{\sim}\mathcal{N}(\mu,\sigma^2)$ on a real line with unknown mean parameter $\mu$ and known standard deviation $\sigma$. Further assume the data sample has been selected based on the following randomized algorithm, which is a simple version of stability selection algorithm.

\begin{framed}
\noindent \textbf{Input:} data vector $(Y_1,\ldots, Y_n)$, the number of subsamples $m$, subsample size $n_s$, threshold $\tau$, 
\begin{enumerate}[leftmargin=*]
	\item For each $i=1,\ldots, m$:
	\begin{adjustwidth}{1em}{}
		 Draw a subsample $(Y_1^{*i},\ldots, Y_{n_s}^{*i})\in\mathbb{R}^{n_s}$ independently with replacement from the original sample $(Y_1,\ldots, Y_n)$ and check whether 	
	\begin{equation} \label{eq:simple:example:selection}
		\sqrt{n}\bar{Y}^{*i}>\tau,
	\end{equation}
	where $\bar{Y}^{*i}=\frac{1}{n_s}\sum_{j=1}^{n_s} Y_j^{*i}$.  
	\end{adjustwidth}
	\item Select the sample $(Y_1,\ldots, Y_n)$ if more than $q\cdot m$, $q\in(0,1)$ pre-specified, events in \eqref{eq:simple:example:selection} have passed.
\end{enumerate}
\noindent \textbf{Output:} Yes or No (whether the data vector passed or not the selection).
\end{framed}
Assuming we are only interested in discovering the positive effects, we only report a resulting $p$-value and a confidence interval for $\mu$ based on a data sample that passed the above selection, i.e.~the data sample survived the \textit{file drawer effect.}
The goal is to do valid inference for the mean parameter $\mu$ given that our data has been selected based on the presented randomized algorithm.

One can attempt to construct $p$-values and confidence intervals for $\mu$ using the fact that $\sqrt{n}(\bar{Y}-\mu)\sim\mathcal{N}(0,\sigma^2)$, where $\bar{Y}=\frac{1}{n}\sum_{i=1}^n Y_i$ is the sample mean. However, in case the sample has been selected for satisfying some property, $\sqrt{n}(\bar{Y}-\mu)$ is no longer normally distributed so the inference based on this distribution of $\bar{Y}$ is not valid. %To conduct valid inference post-selection, we have to correct for the fact that the data has been selected based on the above algorithm. 

To correct for the fact that the data has been selected based on the above algorithm, we base our inference on the conditional distribution of the test statistic $\bar{Y}$, conditioning on the selection event -- that more than $q\cdot m$ randomized events of the form in \eqref{eq:simple:example:selection} succeeded. Given the test statistic $\bar{Y}=x$ denote with $s(x)$ \textit{the probability of selection}, i.e.~that the data vector $(Y_1,\ldots, Y_n)$ with $\bar{Y}=x$ got selected following the above algorithm. To have valid inference on $\mu$ post-selection, we have to use the conditional density of the test statistic $\bar{Y}$. As discussed, the naive (pre-selective, classical) density of $\bar{Y}$, which is $\phi(x;\mu,\sigma^2/n) = \sqrt{\frac{n}{2\pi\sigma^2}} \exp\left(-n\frac{(x-\mu)^2}{2\sigma^2}\right)$, $x\in\mathbb{R}$, the probability density function of $\mathcal{N}(\mu, \sigma^2/n)$, is not valid post-selection.
\ul{A correct way is to base inference for $\mu$ using the \textit{conditional density} of $\bar{Y}$, which is proportional to}
\begin{equation} \label{eq:simple:example:cond:density}
	\underline{\phi(x;\mu,\sigma^2/n)	\cdot s(x), \;\; x\in\mathbb{R}.}
\end{equation}

Even though it is possible to evaluate the function $s(x)$ for finite $n$, we evaluate it asymptotically as $n$ and $n_s$ tend to infinity with $n/n_s$ fixed. In this asymptotic setting, we have $\sqrt{n_s}(\bar{Y}^{*i}-\bar{Y})\overset{\cdot}{\sim}\mathcal{N}(0,\sigma^2)$ conditional on the data. Hence, the selection event in \eqref{eq:simple:example:selection} is asymptotically given data equivalent to 
\begin{equation*}
	\sqrt{n}\bar{Y}+\omega>\tau, \;\; \omega\sim\mathcal{N}\left(0,\sigma^2\cdot n/n_s\right),
\end{equation*}
with the added randomization $\omega$ independent of the original data. 
Thus, conditional on $\bar{Y}=x$, each of the individual selection events in \eqref{eq:simple:example:selection} happens with probability tending to
\begin{equation} \label{eq:simple:example:ps}
	p_s(x)=\mathbb{P}\left\{\sqrt{n}\bar{Y}^{*i}>\tau\Big|\bar{Y}=x\right\}\overset{\cdot}{\sim} \bar{\Phi}\left(\frac{\tau-\sqrt{n}x}{\sigma\sqrt{n/n_s}}\right),
\end{equation}
where $\bar{\Phi}(\cdot)$ denotes the survival function of the standard normal distribution. Given $\bar{Y}=x$, the number of events out of $m$ that pass selection in \eqref{eq:simple:example:selection} is $B\sim\mathcal{B}(m, p_s(x))$, a binomial random variable with the number of trials $m$ and the probability of success $p_s(x)$. Thus, the probability of the having more than $q\cdot m$ successes becomes
\begin{equation} \label{eq:simple:example:binomial}
	s(x)=\mathbb{P}\{B\geq q\cdot m\}.
\end{equation}
Using the asymptotic formula for $p_s(x)$ from \eqref{eq:simple:example:ps}, $s(x)$ has an explicit asymptotic formula.

Given we know $s(x)$, $x\in\mathbb{R}$, asymptotically we use the unnormalized conditional density in \eqref{eq:simple:example:cond:density} to get a numerical approximate of the corresponding normalized density. Denote the normalized conditional density as $f_{\mu}(x)$ and a grid of $x$ values as $x_1,\ldots, x_G$, where $G$ is the grid size. To get a numerical approximation of $f_{\mu}(x)$, we use a one-dimensional discrete exponential family approximation 
\begin{equation*}
	\tilde{f}_{\mu}(x)=\sum_{g=1}^G \exp(\mu\cdot x_g-\Lambda(\mu))\cdot w_g\cdot\mathbb{I}_{x=x_g},	
\end{equation*}
where $w_g=\exp\left(-\frac{nx_g^2}{2\sigma^2}\right)\cdot s(x_g)$ and $\Lambda(\mu)=\log\left(\sum_{g=1}^G w_g\cdot e^{\mu\cdot x_g}\right)$.
This numerical approximation gives us an approximate normalized conditional density of $\bar{Y}$ after selection. This allows testing test $H_0:\mu=\mu_0$ post-selection for any value $\mu_0$. Inverting this test enables the construction of the confidence interval for $\mu$.

Based on this, so called ``simple example,'' we see that the crucial object in constructing conditional density becomes the selection probability given data, i.e.~the function $s(x)$, which we could evaluate explicitly in the presented problem. In more complicated problems, however, we do not have a good way of evaluating $s(x)$ because we either do not have a formula for the individual probability function $p_s(x)$ or we cannot even write $s(x)$ as a survival function of a Binomial distribution. As we will see in Section \ref{sec:learning}, we propose a way to estimate $s(x)$ (or $p_s(x)$) after a possibly randomized black box model selection procedure.

%%%%%%%%%%%%%%%%%%%%%%%%%%%%%%%%%%%%%%%%%%%%%%%%%%%%
%%%%%%%%%%%%%%%%%%%%%%%%%%%%%%%%%%%%%%%%%%%%%%%%%%%%

%---------------------------------------------------
\section{Regression problem setup} \label{sec:regression}
%---------------------------------------------------

Given the fixed design matrix $X\in\mathbb{R}^{n\times p}$ and the response $y\in\mathbb{R}^n$, assume our data follows a linear model
\begin{equation*}
	y=X\beta+\epsilon, \;\;\epsilon\sim\mathcal{N}(0,\sigma^2I_n),	
\end{equation*}
for an unknown vector $\beta\in\mathbb{R}^p$ and an unknown noise variance $\sigma^2$. For simplicity of exposition, we assume $p<n$. 
We run a possibly randomized variable selection algorithm to select a subset $\hat{E}(X^\top y)=E\subset\{1,\ldots,p\}$ of the predictors. We assume the selection procedure $\hat{E}(X^\top y)$ depends on the data vector $D=X^\top y$ and possibly on an independent randomization. In case the model selection procedure is non-randomized, i.e.~there is no independent randomization and the selection algorithm results depend on $X^\top y$ only, the setup in this section still applies.
\ul{After looking at the selection result, the set $E$, the goal is to go valid post-selection inference for each of the parameters $\beta_j$, $j\in E$, corresponding to the observed set $E$.} Later in this section, we discuss the extensions to high-dimensional settings (see Remark \ref{remark:high:dim}) and conducting inference for a different set of selected parameters (see Remark \ref{remark:targets}).

In case $E$ was fixed and not chosen based on the data, inference for $\beta_j$, $j\in E$ would be based on the (pre-selection) normality of the least squares estimate $\bar{\beta}_j-\beta_j\sim\mathcal{N}(0,\sigma_j^2)$, where $\bar{\beta}_j=e_j^\top (X^\top X)^{-1}X^\top y$, $\sigma_j^2=\sigma^2\cdot (X^\top X)^{-1}_{j,j}$ and $\{e_j\in\mathbb{R}^p: j=1,\ldots,p\}$ corresponds to the standard orthogonal basis in $\mathbb{R}^p$. In case $E$ is data-dependent, the normality of $\bar{\beta}_j$ no longer holds -- using the naive $p$-values, constructed based on the normality of this test statistic, will have inflated type I error and the respective confidence intervals will under-cover the true parameter $\beta_j$.

In order to account for the fact that $E$ was selected based on data and to do valid inference on the parameter $\beta_j$, $j\in E$, we modify the normal distribution of our test statistic $\bar{\beta}_j$ by conditioning its (pre-selection) normal distribution on the event that the coordinate $j$ was selected: $j\in\hat{E}(X^\top y)$. Since the conditional distribution of $\bar{\beta}_j\big|j\in\hat{E}(X^\top y)$ depends on not only the parameter of interest $\beta_j$ but also on other parameters, we further condition out the nuisance parameters. 
Since $X^\top y$ and $\bar{\beta}_j$ are jointly Gaussian in case $E$ is fixed, i.e. 
\begin{equation}
	\begin{pmatrix}
		X^\top y \\ \bar{\beta}_j
	\end{pmatrix} \sim \mathcal{N}\left(\begin{pmatrix} X^\top X\beta \\ \beta_j \end{pmatrix}, \begin{pmatrix} \sigma^2\cdot X^\top X & \sigma^2\cdot e_j \\ \sigma^2\cdot e_j^\top & \sigma_j^2 \end{pmatrix} \right)
\end{equation}
we have that $N_j=X^\top y-e_j\frac{\sigma^2}{\sigma_j^2}\bar{\beta}_j$ is independent of $\bar{\beta}_j$. Thus, we can decompose $X^\top y=N_j+d_j\bar{\beta}_j$, where direction $d_j=e_j\frac{\sigma^2}{\sigma_j^2}$, into two independent normal vectors $\bar{\beta}_j$ and $N_j$. To conduct inference (construct $p$-values and confidence intervals), we further condition on $N_j$, so that \ul{the final distribution of $\bar{\beta}_j$ for inference on $\beta_j$ becomes the normal distribution $\mathcal{N}(\beta_j,\sigma_j^2)$ of $\bar{\beta}_j$ conditional on $j\in \hat{E}(N_j+d_j\bar{\beta}_j)$ and $N_j=N_j^{obs}$ fixed at its observed value.}
The probability density of this conditional distribution is 
\begin{equation}
\begin{aligned}
	&f_j(x) =\mathbb{P}\left\{\bar{\beta}_j=x \Big| j\in \hat{E}(N_j+d_j\bar{\beta}_j), N_j=N_j^{obs}\right\} \\
	&=\frac{\mathbb{P}\left\{j\in \hat{E}(N_j^{obs}+d_j\bar{\beta}_j)\Big| \bar{\beta}_j=x\right\}\cdot\mathbb{P}_{\bar{\beta}_j}\left\{\bar{\beta}_j=x\right\}}{\mathbb{P}\left\{j\in \hat{E}(N_j^{obs}+d_j\bar{\beta}_j)\right\}},
\end{aligned}
\end{equation}
where the probabilities above are with respect to $\bar{\beta}_j\sim\mathcal{N}(\beta_j,\sigma_j^2)$ and possible independent randomization.
\ul{The crucial quantity is the probability of $j$-th coordinate being active given data, denoted as}
\begin{equation} \label{eq:sel:prob:full}
	\underline{s_j(x)=\mathbb{P}\left\{j\in \hat{E}(N_j^{obs}+d_jx)\right\}, \;\;x\in\mathbb{R},}
\end{equation}
where the probability is only with respect to the added randomization, fixing the nuisance statistic at its observed value and the test statistic value at $\bar{\beta}_j=x$. In case there is no added randomization but the selection procedure is a deterministic function of $D$, $s_j(x)$ becomes an indicator function and the method still applies.
We refer to the function $s_j(x)$ as the (conditional) selection probability given data throughout the text.
Finally, the conditional density $f_j(x)$, $x\in\mathbb{R}$, of $\bar{\beta}_j$ used for inference becomes proportional to
\begin{equation}
	\phi(x; \beta_j,\sigma_j^2)\cdot s_j(x), 
\end{equation}
where $\phi(x;\beta_j,\sigma_j^2)$ is the density of $\mathcal{N}(\beta_j,\sigma_j^2)$. Since we do not know the function $s_j(x)$ in most cases, this paper proposes estimating it.

\begin{remark}\textbf{(Different parameter targets)} \label{remark:targets}
We defined the data-dependent coefficients of interest to be the selected coordinates of parameter vector $\beta$, which is defined with respect to the original data generating distribution. Hence in this case to conduct inference on $\beta_j$ for a single $j\in E$, it is sufficient to condition on only a particular coordinate $j$ being active, without conditioning on the whole active set \cite{liu2018more}. To construct $p$-values and confidence intervals for the different kind of data-dependent parameters, the so called ``partial targets,'' we have to condition on observing the whole active set since defining each coordinate of the partial target requires knowing the whole active set $E$. The details of the proposed method in the case of partial targets are in Appendix \ref{sec:partial:targets}.
\end{remark}

\begin{remark}\textbf{(High-dimensional setting)} \label{remark:high:dim}
We restricted our discussion to low-dimensional $p<n$ setting in order to have a simple expression for the test statistic $\bar{\beta}_j$. In case $p>n$, this test statistic is no longer defined since $X^\top X$ is not invertible. In the high-dimensional settings, we use the debiased Lasso test statistic introduced in \cite{javanmard2014confidence} instead of the least squares estimate $\bar{\beta}$. The whole conditioning argument still applies as described -- we omit the details.
\end{remark}

%%%%%%%%%%%%%%%%%%%%%%%%%%%%%%%%%%%%%%%%%%%%%%%%%%%%
%%%%%%%%%%%%%%%%%%%%%%%%%%%%%%%%%%%%%%%%%%%%%%%%%%%%

%---------------------------------------------------
\section{Learning selection probabilities} \label{sec:learning}
%---------------------------------------------------

We explain how the unknown function $s_j(x)$, $x\in\mathbb{R}$, is estimated for each $j$ using a statistical learning model. We describe the procedure for a single $j\in E$ -- the procedure is then repeated to estimate all $s_j(x)$, $j\in E$. 

On a high level, we learn the selection probability function $s_j(x)$ by perturbing the data along the line and re-running the original black box  algorithm for the new data vector. The details are as follows. 
We generate values $Z_1,\ldots, Z_B$ from a probability distribution on a real line (see Remark \ref{remark:covariates} for the details on the distribution generating these covariates). Then we re-run a possibly randomized selection algorithm using data vector $N_j^{obs}+d_jZ_b$. In case the model selection algorithm is non-randomized, we re-run the model selection algorithm on $N_j^{obs}+d_jZ_b$ without adding any randomization to it. For the notational purposes, we assume the model selection algorithm is randomized.

We denote as $W_b\in\{0,1\}$ the indicator variable, taking value one in case variable $j$ is selected using the perturbed data vector $N_j^{obs}+d_jZ_b$ and an independent randomization and zero otherwise:
\begin{equation}
	W_b = \mathbb{I}_{\left\{j\in\hat{E}(N_j^{obs}+d_jZ_b)\right\}}.
\end{equation}

Our data used for learning now consists of $\{(Z_b,W_b)\}_{b=1}^B$ and the goal is to learn the probability of observing positive outcome
\begin{equation}
	s_j(x)=\mathbb{P}\{W=1\big|Z=x\}	,
\end{equation}
where $(Z,W)$ comes from a probability distribution underlying data generating mechanism of each of the data samples $\{(Z_b,W_b)\}_{b=1}^B$.
This problem now looks more like a standard statistical learning or estimation problem, where we estimate the probabilities of success given binary response and one-dimensional covariates.

\begin{remark} \textbf{(Generating covariates)} \label{remark:covariates}	
Although we could generate covariates $Z_1,\ldots, Z_B$ deterministically, in practice we draw them from a normal distribution centered at the observed value of $\bar{\beta}_j$ and with scaled estimated standard deviation $\hat{\sigma}_j^2$, where the scaling is chosen separately for each point $Z_b$ uniformly at random from (0.5, 1, 1.5, 2).
\end{remark}

To estimate $s_j(x)$ based on the generated data $\{(Z_b,W_b)\}_{b=1}^B$, we propose the following generalized linear models (GLM) with two different link types.
\begin{enumerate}[leftmargin=*]
	\item \textbf{GLM with probit link} models the selection probability function $s_j(x)$ as
	\begin{equation*}
		s_j(x)\approx\Phi(\gamma^\top ns(x, 10)),	
	\end{equation*}
	where $ns(x,10)$ denotes a natural spline with 10 degrees of freedom and $\gamma$ is an unknown vector parameter.
	 
	\item \textbf{GLM with logit link} models $s_j(x)$ as
	\begin{equation*}
		s_j(x)\approx\frac{\exp(\gamma^\top ns(x,10))}{1+\exp(\gamma^\top ns(x,10))}.	
	\end{equation*}
\end{enumerate} 

We fit each of the models on the generated data $\{(W_b,Z_b)\}_{b=1}^B$ to estimate unknown parameter $\gamma$, giving an estimate $\hat{s}_j(x)$ of the whole function $s_j(x)$, $x\in\mathbb{R}$. We evaluate and compare these two methods on several simulated examples in the sections that follow.

\begin{remark} \textbf{(Estimation target)} \label{remark:est:target}
In case the selection probabilities $s_j(x)$ can be represented as the Binomial $\mathcal{B}(m, p_{s_j}(x))$ survival function \eqref{eq:simple:example:binomial}, we can also apply the above estimation procedure to estimate the individual probability of success $p_{s_j}(x)$ and not directly $s_j(x)$. This involves re-running the selection procedure once and not $m$ times for each $b=1,\ldots, B$. Having estimated $p_{s_j}(x)$, the selection probability $s_j(x)$ is estimated using the survival function of the corresponding Binomial distribution. We use this approach in Section \ref{sec:cv}.
\end{remark}

%%%%%%%%%%%%%%%%%%%%%%%%%%%%%%%%%%%%%%%%%%%%%%%%%%%%%%
%%%%%%%%%%%%%%%%%%%%%%%%%%%%%%%%%%%%%%%%%%%%%%%%%%%%%%

%-----------------------------------------------------
\section{Simple example revisited} \label{sec:simple:example:revisited}
%-----------------------------------------------------

Even though in the case of the simple example from Section \ref{sec:simple:example} there is an explicit (asymptotic) formula for $s(x)$, we show here the estimation procedure for $s(x)$ in a simple setting.

The procedure for generating data $\{(Z_b, W_b)\}_{b=1}^B$ is as follows.
\begin{framed}
\noindent For each $b=1,\ldots, B$:
\begin{enumerate}[leftmargin=*]
	\item Draw $Z_b$ from a distribution as mentioned in Remark \ref{remark:covariates}.
	\item For each $i=1,\ldots,m$: 
	\begin{adjustwidth}{1em}{}
	Draw independent randomization $\omega_{b,i}\sim\mathcal{N}(0,\sigma^2\cdot n/n_s)$ and check whether $\sqrt{n}Z_b+\omega_{b,i}>\tau$.
	\end{adjustwidth}
 	\item Define variable $W_b$ to be one if the above selection events passed for more than $q\cdot m$ events.
\end{enumerate}
\end{framed}
Given the generated data, the learning methods from Section \ref{sec:learning} give us the estimate of $s(x)$, $x\in\mathbb{R}$.

%-------------------------------------------------
\subsection{Simulation results}
%-------------------------------------------------

We show the simulation results in the setting with the following parameters: $(n=100, n_s=50, m=20, q=0.5, \sigma=1, \mu=0, B=10000, nsims=1000)$, where $nsims$ denotes the number of repeated simulations. Each of the $nsims$ simulated data vectors has passed the selection event.

\begin{figure}[h!]
  \centering
    \includegraphics[width=0.6\textwidth]{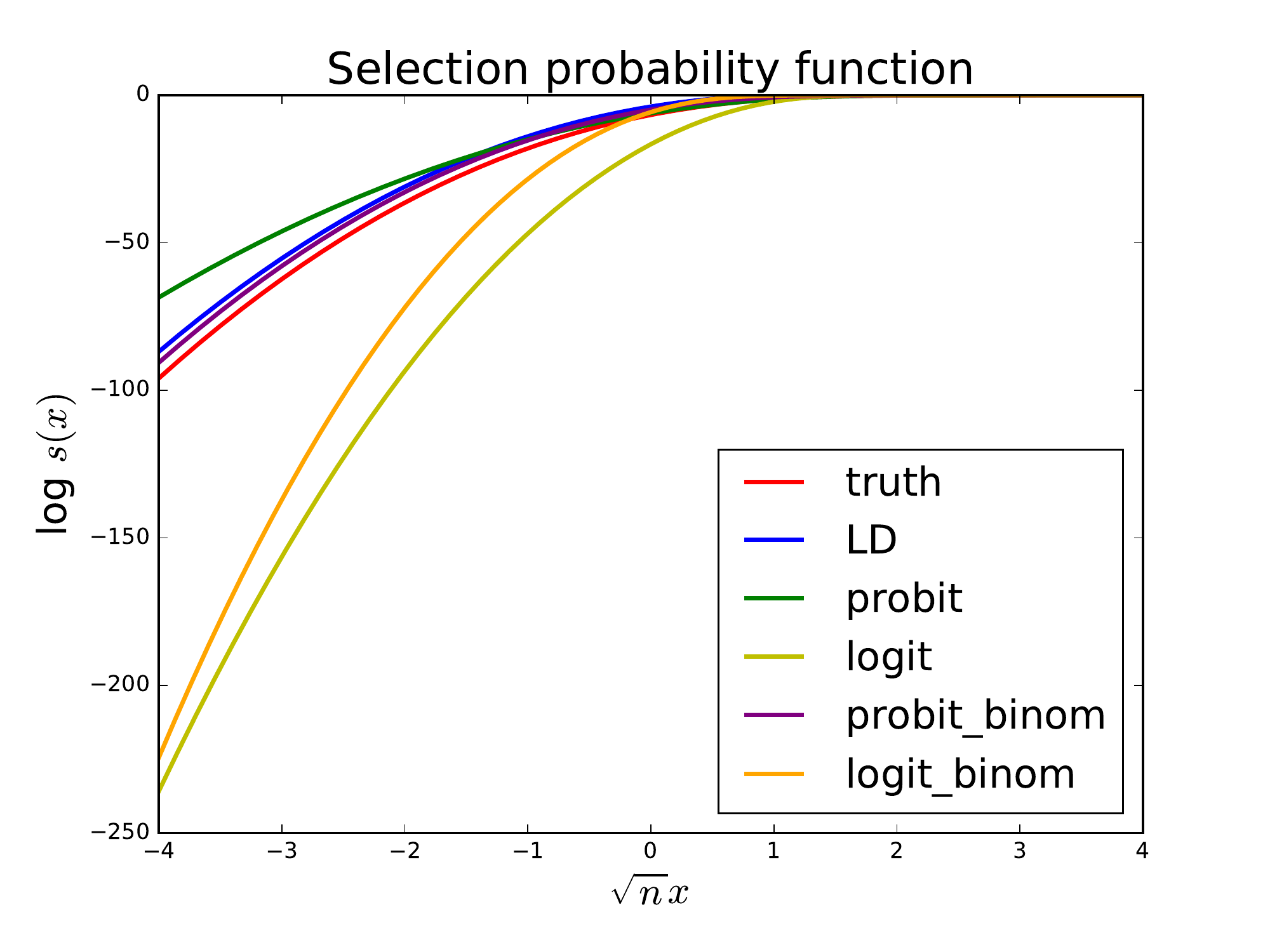}
    \caption{The logarithm of the selection probability function $s(x)$ (true asymptotic), along with its large deviation (LD) approximation and its estimates via probit and logit fit in the simple example.}
    \label{fig:simple:example:sel:prob}
\end{figure}

In Figure \ref{fig:simple:example:sel:prob}, we show the true asymptotic function $s(x)$, $x\in\mathbb{R}$, computed using the formula in \eqref{eq:simple:example:binomial}, along with the large deviation (LD) approximation of binomial survival function in \eqref{eq:simple:example:binomial}. We further add the estimates of $s(x)$ using probit and logit fit, each using $B$ generated samples. The figure shows that the estimated functions follow the true function closely while LD approximation is slightly worse.

In Figure \ref{fig:simple:example:pivots}, we show the empirical CDFs of the selective $p$-values for testing $H_0:\mu=0$ using each of the methods above, plus the empirical CDF of the naive $p$-values that ignore selection took place. The results show that the $p$-values using true asymptotic $s(x)$ or any of the estimates are uniform, hence valid, while the ones constructed using LD approximation and the naive ones are much further from uniform.

\begin{minipage}{\textwidth}
\begin{minipage}[b]{0.55\textwidth}
%\begin{figure}[h!]
  \centering
    \includegraphics[width=\textwidth]{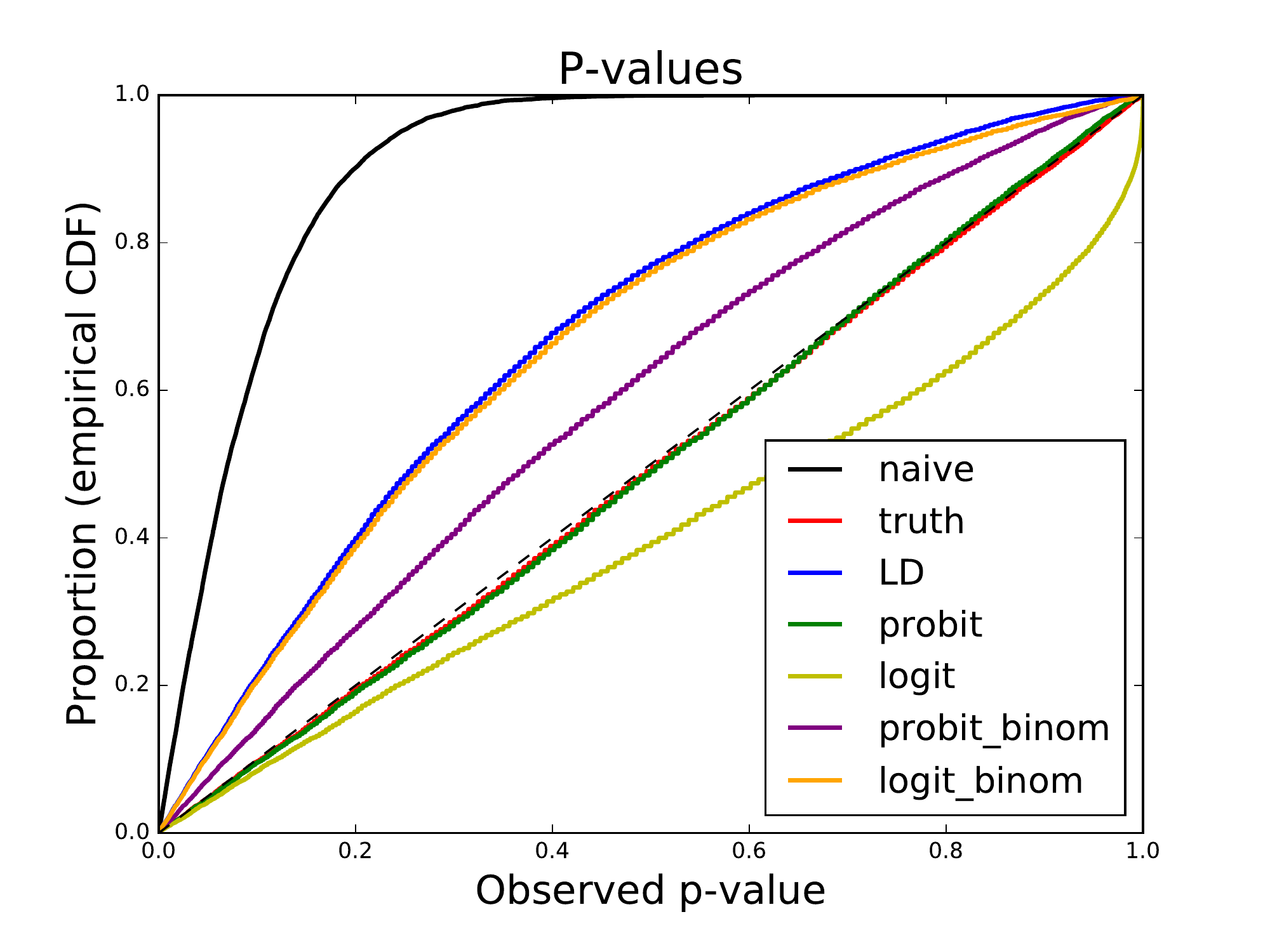}
   \captionof{figure}{Emprirical CDFs of the selective $p$-values using various methods of obtaining $s(x)$ plus the empirical CDF of the naive $p$-values for testing $H_0:\mu=0$ in the simple example.}
    \label{fig:simple:example:pivots}
%\end{figure}
\end{minipage}
  \hfill
 \begin{minipage}[b]{0.43\textwidth}
 \centering
  \begin{tabular}{cc}\hline
      Methods & Coverages (in \% )\\ \hline
      naive & 80.6 \\
      truth & 95.3 \\
      LD & 92.9 \\
      probit & 94.9 \\
      logit & 83.2 \\
      probit binom & 95.4 \\ 
      logit binom & 93.2 \\ \hline
      \end{tabular}
      \captionof{table}{Empirical coverages of the selective confidence intervals with target coverage of 95\% using various methods of obtaining $s(x)$ plus the empirical CDF of the naive confidence intervals $\mu$.}	
\end{minipage}
\end{minipage}

%%%%%%%%%%%%%%%%%%%%%%%%%%%%%%%%%%%%%%%%%%%%%%%%%%%%
%%%%%%%%%%%%%%%%%%%%%%%%%%%%%%%%%%%%%%%%%%%%%%%%%%%%

%---------------------------------------------------
\section{Stability selection with Lasso} \label{sec:stability:selection}
%---------------------------------------------------

Given the inference via estimation method described in Section \ref{sec:regression} and Section \ref{sec:learning}, we apply the method to learn the selection probability function for a specific black box algorithm -- stability selection of \cite{meinshausen2010stability}.

Given the data $X$ and $y$ as in Section \ref{sec:regression}, we explain the stability selection algorithm which is in this case used to enhance the Lasso procedure. Standard Lasso runs the following objective
\begin{equation} \label{eq:lasso}
	\hat{\beta}_{\lambda}(X^\top y,X^\top X)=\textnormal{arg}\underset{\beta\in\mathbb{R}^p}{\min}\left\{\frac{1}{2}\|y-X\beta\|_2^2+\lambda\|\beta\|_1\right\}
\end{equation}
for a pre-specified penalty level $\lambda$, where we also use $\hat{\beta}_{\lambda}$ for the Lasso solution for short. The selected variables are then the ones for which the corresponding $\hat{\beta}_{\lambda}$ takes a non-zero value: $E_\lambda=\left\{j:\hat{\beta}_{\lambda,j}\neq 0\right\}$. Since the set $E_{\lambda}$ is sensitive to the choice of $\lambda$, \cite{meinshausen2010stability} suggest the following stability selection algorithm to get a more robust selection set.

\begin{framed}
\noindent \textbf{Input:} data $(X,y)$, the number of subsamples $m$, threshold $q$. 
\begin{enumerate}[leftmargin=*]
\item For $i=1,\ldots,m$:
\begin{enumerate}
\item Subsample the data randomly with replacement to contain 50\% of the original data size. The subsample is denoted as $(X^{*i}, y^{*i})\in\mathbb{R}^{\lfloor n/2\rfloor \times p}\times\mathbb{R}^{\lfloor n/2\rfloor}$.

\item For each subsample $(X^{*i}, y^{*i})$, re-run the Lasso for $\lambda$ values in  $[\lambda_{\min},\lambda_{\max}]$. Value $\lambda_{\min}$ is such that the Lasso on $(X,y)$ selects at most $\sqrt{0.8p}$ variables and $\lambda_{\max}$ is chosen so that the Lasso on the same dataset does not select any variable. For each of the subsamples and each $\lambda$, the Lasso gives a new set $E_{\lambda}^{*i}$.
\end{enumerate}

\item For each $\lambda$ and each predictor $j$, denote the proportion of sets $E^{*i}_{\lambda}$, $i=1,\ldots,m$ containing the predictor $j$:
\begin{equation*}
	q_{\lambda, j} = \frac{1}{m}\sum_{i=1}^m \mathbb{I}_{\{j\in E_{\lambda}^{*i}\}}.
\end{equation*}

\item The final set is taken to consist of all the variables $j$ for which there exists $\lambda$ so that for that $\lambda$ value predictor $j$ shows above a certain number of times in $E_{\lambda}^{*i}$ sets, $i=1,\ldots,m$:
\begin{equation*}
	E = \left\{j: \exists \lambda \textnormal{ s.t. }q_{\lambda,j} \geq q\right\},
\end{equation*}
where $q\in(0,1)$ is a pre-specified threshold.
\end{enumerate}
\textbf{Output:} set $E$.
\end{framed}
Given the final selected set $E$, our method provides valid inference for the predictors in $E$. We achieve that by minimally conditioning on each of the final active variables in $E$, without any conditioning on the intermediate sets $E_{\lambda}^{*i}$.

\begin{comment}
The presented stability selection algorithm is a complex randomized selection algorithm with no explicit way to put the constraints on the data and randomization so that the resulting selection set is always the observed selection set $E$. We take this algorithm as an instance of our black box algorithm we use for selection. In the inference step, we run the stability selection algorithm many times for perturbed data values and independent randomization in order to learn the selection probabilities.
\end{comment}

Before writing down the full proposed algorithm for conducting selective inference after selecting $E$, we should be precise about how exactly we perturb the data and choose the randomization.
Since we are directly perturbing $X^\top y$ to get a new data vector $D_j(x)=N_j^{obs}+d_jx$, we do not have a new $X$ and $y$ to subsample from when re-running the algorithm. We turn to the asymptotics to solve this problem as in the simple example in Section \ref{sec:simple:example}. Since $2{X^*}^\top y^*-X^\top y\sim\mathcal{N}(0, 2\cdot\sigma^2\cdot X^\top X)$, we have that the asymptotic equivalent of running multiple Lasso algorithms $\hat{\beta}_{\lambda}({X^*}^\top y^*, {X^*}^\top X^*)$ on the subsamples is equivalent to running the Lasso on $\hat{\beta}_{\lambda}(X^\top y/2+\omega, X^\top X/2)$, where $\omega\sim\mathcal{N}(0, 0.5\cdot\sigma^2\cdot X^\top X)$.

In the case the stability selection is our black box randomized algorithm, the algorithm for generating the data $\{(Z_b,W_b)\}_{b=1}^B$, is as follows.
\begin{framed}
\noindent For each $b=1,\ldots, B$:
\begin{enumerate}[leftmargin=*]
	\item Draw $Z_b$ from a probability distribution as mentioned in Remark \ref{remark:covariates}.
	\item \label{item:individual:ss} (Stability selection) Recompute $D_j(Z_b)$ and draw $m$ independent randomizations $\omega_{b,i}\sim\mathcal{N}(0, 0.5\cdot\sigma^2\cdot X^\top X)$, $i=1,\ldots m$. For each $i=1,\ldots, m$: run $m$ Lasso paths to get $\hat{\beta}_{\lambda}(D_j(Z_b)/2+\omega_{b,i}, 2\cdot X^\top X)$ and their corresponding supports $E_{\lambda}^{b, *i}$ across $\lambda$ values.
		\item Compute the proportion of sets $E_{\lambda}^{b, *i}$ containing predictor $j$:
		\begin{equation*}
			q_{\lambda, j}^b = \frac{1}{m}\sum_{i=1}^m\mathbb{I}_{\{j\in E_{\lambda}^{b,*i}\}}.
		\end{equation*}

		\item Define variable $W_b$ to be one if there exists $\lambda$ for which $q_{\lambda,j}^b$ is above level $q$. 
\end{enumerate}
\end{framed}
%Step \eqref{item:individual:ss} of the algorithm above corresponds to an individual stability selection algorithm in which one re-runs the Lasso path for each of $m$ subsamples. 
After running the whole algorithm, the resulting data $\{(Z_b,W_b)\}_{b=1}^B$ is then used in estimating the function $s_j(x)$ using the learning techniques in Section \ref{sec:learning}.

%----------------------------------------
\subsection{Simulation results} \label{sec:ss:simulation}
%----------------------------------------

We combine the algorithm in Section \ref{sec:stability:selection} to generate the data $\{(Z_b,W_b)\}_{b=1}^B$ and the learning method in Section \ref{sec:learning} to fit two GLM models on this data, each giving an estimate of $s_j(x)$, $x\in\mathbb{R}$. We then repeat the process for all variables $j$ in the final set $E$.

The rows of $X$ have been generated as i.i.d.~vectors from $\mathcal{N}(0,\Sigma)$, where $\Sigma$ follows autoregressive $AR(\rho)$ model. The regression coefficients vector $\beta=\beta^{true}$ is taken to have sparsity $s$ with non-zero coefficients equal to $\pm b$ with equal probability. 
In the simulations, we take the following simulation parameters
$(n=200, p=100, \rho=0.1, \sigma=2, s=20, b=3, m=5, q=0.6, B=1000, nsims=50)$. Figure \ref{fig:ss:pivots} shows the empirical CDF of the constructed selective pivots for testing the $H_0:\beta_j=\beta_j^{true}$, $j\in E$, across $nsims$ simulated $(X,y,\beta)$ settings. We see that the selective pivots are roughly uniform when either probit or logit fit is used to estimate the selection probabilities. The figure further includes the empirical CDF of the naive pivots for testing the same hypotheses, empirically showing these are not valid. Selective confidence intervals constructed with the targeted level of 90\% using probit and logit estimates have 89\% and 90\% coverage, respectively, while naive confidence intervals have 64\% coverage for the same target parameters.

\begin{figure}[h!]
  \centering
    \includegraphics[width=0.6\textwidth]{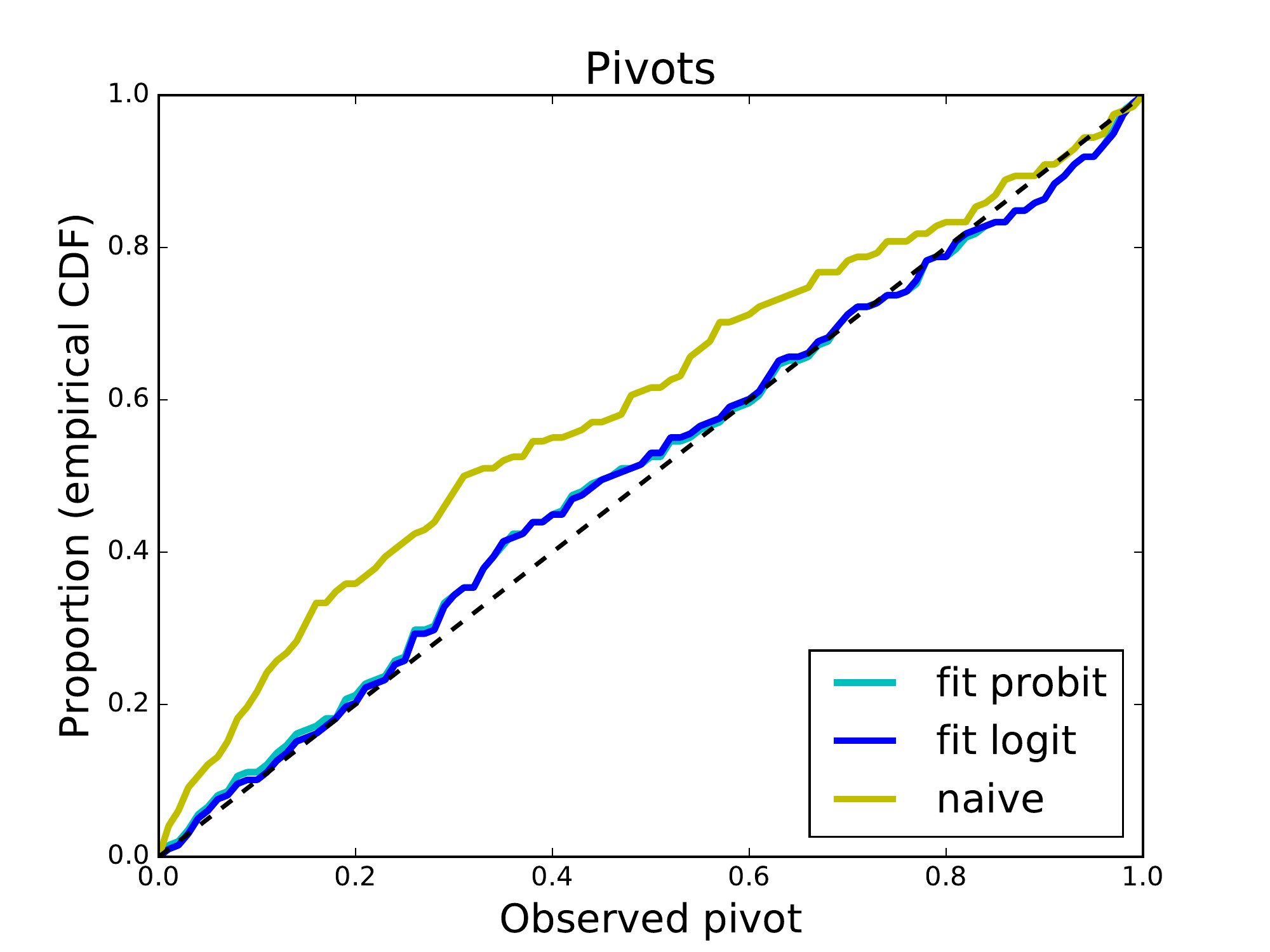}
    \caption{Empirical CDFs of selective inference pivots using probit and logit estimates of the selection probabilities and the emprirical CDF of naive pivots after stability selection algorithm. }
    \label{fig:ss:pivots}
\end{figure}

%---------------------------------------
\section{Multiple cross-validation} \label{sec:cv}
%---------------------------------------

Cross-validation is used to choose the penalty level $\lambda$ in the Lasso objective \eqref{eq:lasso}. The data $(X,y)$ is randomly split into $K$ folds and for each $\lambda$ across the grid and each fold $k=1,\ldots, K$ the Lasso estimator is fit on the leftover folds. The test error is then estimated on fold $k$. For each $\lambda$ the test errors are averaged across folds and the resulting $\lambda^{CV}$ is chosen to minimize the resulting cross-validated error. Running the Lasso procedure on the original data with the penalty level $\lambda^{CV}$ gives a set of selected variables $\textnormal{supp}(\hat{\beta}_{\lambda^{CV}}(X^\top y, X^\top X))=E_{\lambda^{CV}}$.

To further assess the robustness of the procedure, we repeat the above CV algorithm on multiple data subsamples, each with a different random split. Running the CV procedure $m$ times on the data subsamples gives a sequence of cross-validated penalty levels $\lambda^{CV}_1, \ldots, \lambda^{CV}_m$ and their corresponding selection sets $E_{\lambda^{CV}_1},\ldots, E_{\lambda^{CV}_m}$. As a final set $E$, we choose the set consisting of variables appearing in more than $q\cdot m$ sets, with $q\in(0,1)$ pre-specified fraction.

We apply the estimation framework developed in this work to estimate the conditional selection probabilities given data -- function $s_j(x)$ for selected $j\in E$. See also Remark \ref{remark:est:target} for details related to this example.

%-------------------------------------------
\subsection{Simulation results}
%-------------------------------------------

The setting is the same as in Section \ref{sec:ss:simulation} with the following simulation parameters $(n=200, p=50, \rho=0., \sigma=2, s=10, b=2, m=3, q=2/3, B=100, nsims=100)$.

\begin{figure}[h!]
  \centering
  \includegraphics[width=0.6\textwidth]{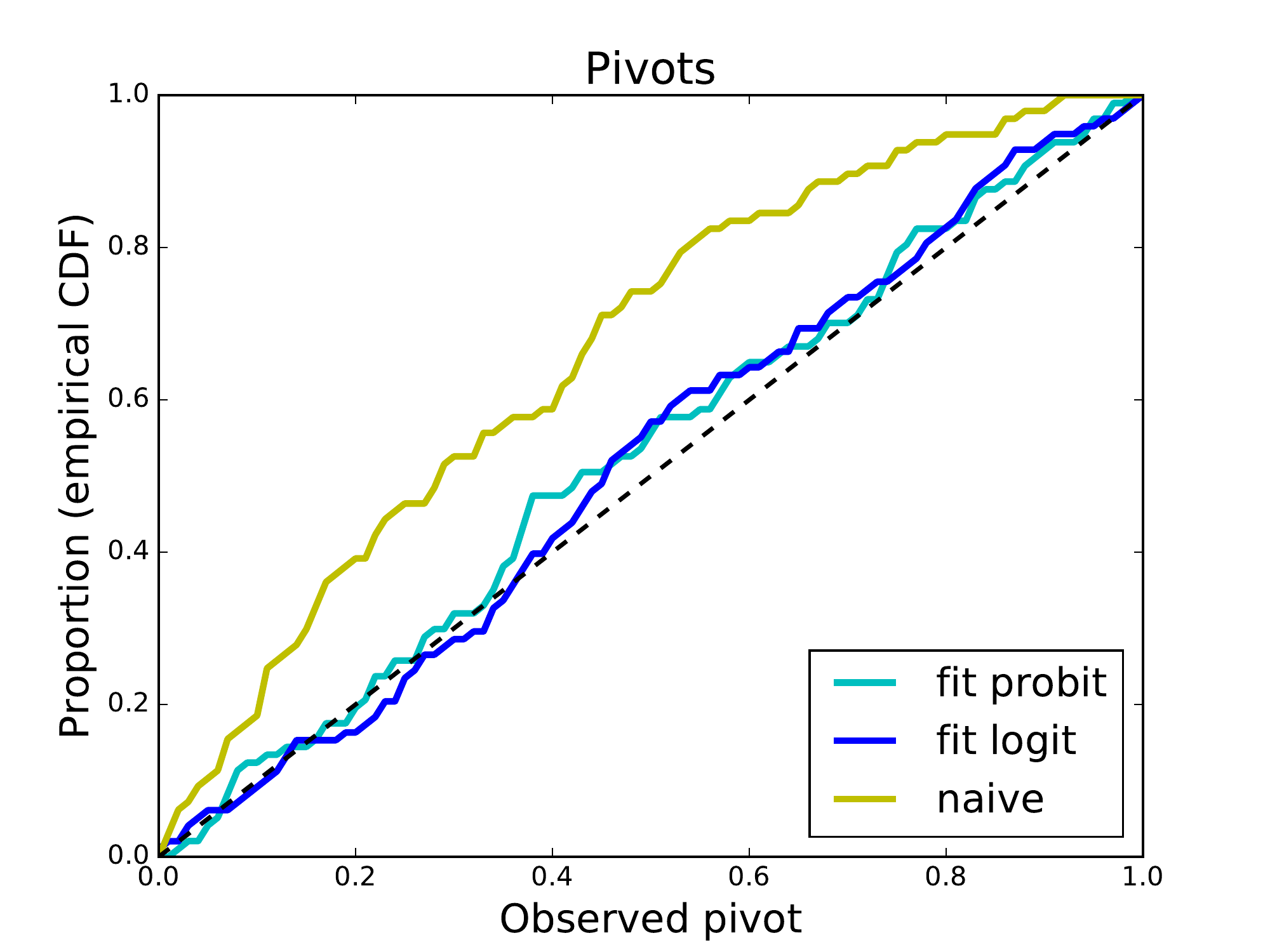}
  \caption{Selective inference pivots using probit and logit estimates of the selection probabilities and the naive pivots after multiple CV procedures.}
    \label{fig:ss:pivots}
\end{figure}

%---------------------------------------------------
\section{Discussion and future research} \label{sec:discussion}
%---------------------------------------------------

In this work, we reframed the selective inference problem as a pure estimation problem, allowing us to develop a method for conducting inference after black box variable selection algorithms. In terms of dimensions of problem, for moderate computational costs we can address problems in dimensions in the hundreds, higher than those of POSI \cite{berk2013valid}, our closest competitor in terms of generality of selection. It is worth noting, however, that generating the data $\{(Z_b, W_b)\}_{b=1}^B$ used in estimation is highly parallelizable since all the points are created independently.

To further improve on computation, a promising future direction includes applying \textit{active learning} to our estimation problem.  Active learning algorithms attempt to minimize the amount of data required to learn a classifier by selecting what data points to label and include in model training \cite{active-learning}. Active learning is particularly useful where the cost of labeling is high. In the current context, the cost of labeling corresponds to evaluating or sampling $s_j(x)$. Active learning heuristics could be used to select points $x$ near where $s_j(x)\approx 0.5$ so as to provide high resolution near regions of interest.

Future research also includes the following important questions: theoretical results on how well one needs to estimate the function $s_j(x)$ so that the inference using the conditional density with estimated $s_j(x)$ is valid and changing the learning model for estimating the function $s_j(x)$ to include more complicated statistical models, e.g.~neural networks.

\begin{comment}
\begin{enumerate}[leftmargin=*]
	\item Theoretical results on how well one needs to estimate the function $s_j(x)$ so that the inference using the conditional density with estimated $s_j(x)$ is valid would be crucial for not only having guarantees but also practical guidance. The questions include determining the optimal sample size $B$, the underlying distribution of the sampled covariates $Z$ used in learning and the best statistical model explaining $Y$ in terms of $Z$ for learning $s_j(x)$.
	
	\item Practically, changing the learning model for estimating the function $s_j(x)$ to include more complicated statistical models, e.g.~neural networks, instead of GLM would give our method the ability to learn more complicated selection probability functions, arising from even more complex randomized selection algorithms.
	 
	\item Developing computationally efficient algorithm for learning $s_j(x)$ fast would be crucial for conducting post-selection inference after model free knockoffs mentioned in the introduction and other more complex randomized selection procedures that require more computation. 
\end{enumerate}
\end{comment}

%%%%%%%%%%%%%%%%%%%%%%%%%%%%%%%%%%%%%%%%%%%%%%%%%%%%
%%%%%%%%%%%%%%%%%%%%%%%%%%%%%%%%%%%%%%%%%%%%%%%%%%%%

%---------------------------------------------------
\appendix
\section*{Appendices}
\addcontentsline{toc}{section}{Appendices}
\renewcommand{\thesubsection}{\Alph{subsection}}

\subsection{Partial targets} \label{sec:partial:targets}
%---------------------------------------------------

Following the regression setting in Section \ref{sec:regression}, we change the target of inference so that the inference is no longer done on the coefficients $\beta_j$, $j\in E$, where $E$ is the selected model. Assume now the data follows $y\sim \mathcal{N}(\mu,\sigma^2 I_n)$ and $X$ is fixed. We do not assume $p<n$ here, so the discussion here applies in high dimensions as well. The partial targets are defined as
\begin{equation*}
	\beta_E^*=(X_E^\top X_E)^{-1}X_E^\top \mu,
\end{equation*}
where $X_E$ is the sub-matrix of $X$ consisting of columns of $X$ that are in $E$. The goal here is to make inference on individual coefficients $\beta_{E,j}^*$, $j=1,\ldots, |E|$. Note that in order to define a single coordinate $\beta_{E,j}^*$ one needs to know the whole set $E$ whereas in the setting in Section \ref{sec:regression} one needs to know only whether that single coordinate $j$ is active or not. This difference shows as more conditioning is required for doing inference on the partial targets.

The test statistic used for inference on $\beta_{E,j}^*$ is $\bar{\beta}_{E,j}=e_{E,j}^\top(X_E^\top X_E)^{-1}X_E^\top y$, where $e_E=(e_{E,1},\ldots, e_{E,|E|})$ is the standard basis in $\mathbb{R}^{|E|}$. 
The joint vector $\begin{pmatrix} X^\top y\\ \bar{\beta}_{E,j}\end{pmatrix}$ is normally distributed as
\begin{equation*}
\mathcal{N}\left(\begin{pmatrix} X^\top \mu \\ \beta_{E,j}^* \end{pmatrix}, 
\begin{pmatrix} \sigma^2 X^\top X & \sigma^2 X^\top (X_E^\dagger)^\top e_{E,j} \\ \sigma^2 e_{E,j}^\top X_E^\dagger X & \sigma_{E,j}^2 \end{pmatrix} \right),
\end{equation*}
 where $\sigma_{E,j}^2=\sigma^2\cdot(X_E^\top X_E)^{-1}_{j,j}$ and $X_E^\dagger=(X_E^\top X_E)^{-1}X_E^\top$. The data vector $X^\top y$ decomposes into a sum of two independent variables $N_{E,j}$ and $\bar{\beta}_{E,j}$ as $N_{E,j}+d_{E,j}\bar{\beta}_{E,j}$, where $d_{E,j}= (\sigma^2/\sigma_{E,j}^2)\cdot X^\top (X_E^\dagger)^\top e_{E,j}$ is a fixed vector.

As before, to do valid post-selection inference on data dependent parameter $\beta_{E,j}^*$ we use the conditional density of $\bar{\beta}_{E,j}$  proportional to
\begin{equation} \label{eq:sel:prob:partial}
	\phi(x;\beta_{E,j}^*, \sigma_{E,j}^2)\cdot s_{E,j}(x), \;\; 	x\in\mathbb{R},
\end{equation}
where $s_{E,j}(x)$ is the selection probability given the nuisance statistic $N_{E,j}$ is at its observed value and the test statistic $\bar{\beta}_{E,j}=x$
\begin{equation*}
	s_{E,j}(x)=\mathbb{P}_{\omega}\left\{\hat{E}(N_{E,j}^{obs}+d_{E,j}x,\omega)=E\right\}.	
\end{equation*}
Note that this is the probability of selecting the whole set $E$ and not only the probability of $j$ being active as in \eqref{eq:sel:prob:full}. 

The estimation of the function $s_{E,j}(x)$ for each $j=1,\ldots, |E|$ can be then done analogously to the approach in Section \ref{sec:learning}. This gives an evaluation of the conditional density in \eqref{eq:sel:prob:partial}, enabling us to do both testing and constructing confidence intervals for the selective parameters $\beta_{E,j}^*$.

%%%%%%%%%%%%%%%%%%%%%%%%%%%%%%%%%%%%%%%%%%%%%%%%%%%%%%
%%%%%%%%%%%%%%%%%%%%%%%%%%%%%%%%%%%%%%%%%%%%%%%%%%%%%%

%-----------------------------------------------------
\subsection{General setup} \label{sec:general:setup}
%-----------------------------------------------------

We present a method for conducting selection inference for data-dependent parameter chosen after running looking at the outcome of a black box selection algorithm in a general setting.

Assume the dataset is $S\sim\mathbb{F}$, coming from a data generating distribution $\mathbb{F}$. The black box algorithm $\hat{E}$ takes the data vector $D=D(S)$ and outputs $\hat{E}(D)=E$. The goal is inference for the data-dependent parameter $\theta=\theta(\mathbb{E},E)$ using the original dataset $S$. The parameter depends on the result $E$ of the selection algorithm. 

Assume pre-selection, treating $E$ as fixed, there is a test statistic $T$ that is asymptotically normal around $\theta$, i.e.~$T\overset{\cdot}{\sim}\mathcal{N}(\theta,\Sigma_T)$.
Post-selection distribution of $T$ is then 
\begin{equation*}
	\mathcal{N}(\theta,\Sigma_T) \:\big|\:\hat{E}(D)=E.
\end{equation*}
Further assuming $T$ and $D$ are jointly asymptotically normal so we can decompose 
\begin{equation*}
	D=N+\Sigma_{D,T}\Sigma_T^{-1}T,
\end{equation*}
where $N$ is the statistic corresponding to the nuisance parameters since $N$ and $T$ are asymptotically orthogonal.
Fixing $N$ at its observed value $N^{obs}$, post-selection distribution of $T$ is then 
\begin{equation} \label{eq:post:selection:T}
\mathcal{N}(\theta,\Sigma_T)\:\big|\:\hat{E}(N+\Sigma_{D,T}\Sigma_T^{-1}T)=E, N=N^{obs}.	
\end{equation}
Fixing $N$ at its observed value allows the conditioning event to be written only in terms of $T$ and not in terms of the whole vector $D$. Based on the distribution in \eqref{eq:post:selection:T}, the asymptotic post-selection density of $T$ is
\begin{equation*}
	\phi(x;\theta,\Sigma_T)\cdot s(x), \;\; x\in\mathbb{R}^{\dim(\theta)},
\end{equation*}
where
\begin{equation*}
	s(x)=\mathbb{P}\left\{\hat{E}(N+\Sigma_{D,T}\Sigma_T^{-1}x)=E, N=N^{obs}\right\}
\end{equation*}
is the selection probability given data.

In order to do valid inference on $\theta$ using the asymptotic post-selection density of $T$, we need to know the function $s(x)$. Since we do not have the explicit form of this function, we estimate it. By perturbing the data vector $D$ into $D(x)=N+\Sigma_{D,T}\Sigma_T^{-1}x$, we re-run the black box algorithm on $D(x)$ to get $\hat{E}(D(x)))$ and record whether $\hat{E}(D(x)))$ matches the observed outcome on $E$ or not. The indicators of whether $E$ got selected or not along with features $x$ make up the dataset used in estimating $s(x)$. The rest of procedure is the same as in regression setup.

\newpage

\bibliography{cite}
\bibliographystyle{apalike}

\end{document}